\documentclass[showpacs,epsf,twocolumn]{revtex4-1}
\usepackage{float}
\usepackage{color,graphicx}
\usepackage{hyperref}
\begin{document}

\title{Probing the Fermi Surface of 3D Dirac Semimetal Cd$_{3}$As$_{2}$ through de Haas-van Alphen Technique}

\author{A. Pariari, P. Dutta and P. Mandal}

\affiliation{Saha Institute of Nuclear Physics, 1/AF Bidhannagar, Calcutta 700 064, India}
\date{\today}

\begin{abstract}
We have observed Shubnikov-de Haas  and de Haas-van Alphen effects in the single crystals of three dimensional Dirac semimetal Cd$_{3}$As$_{2}$ upto to 50 K, traceable at field as low as 2 T and 1 T, respectively. The values of Fermi wave vector, Fermi velocity, and effective cyclotron mass of charge carrier,  calculated from both the techniques, are close to each other and match well with the earlier reports.  However, the de Haas-van Alphen effect clearly reflects the existence of two different Fermi surface cross-sections along certain direction and a non-trivial   Berry's phase which is the signature of 3D Dirac Fermion in Cd$_{3}$As$_{2}$.\\

\end{abstract}

\maketitle
Now a days, topology-dependent electronic properties of solids are the subject of considerable research interest in condensed matter physics. Dirac materials such as graphene \cite{castro}, topological insulator \cite{hasan,qi} and the parent compounds of iron based superconductors \cite{rich} whose excitations obey a relativistic Dirac-like equation have been widely studied in recent years. Three dimensional (3D) Dirac semimetal is one of the candidates of topology protected band structure in which semimetal bulk has linearly dispersing excitation  and surface state is the topology protected Fermi arc \cite{ler,hsi,tang,tang1}.
\begin{figure}
\includegraphics[width=0.5\textwidth]{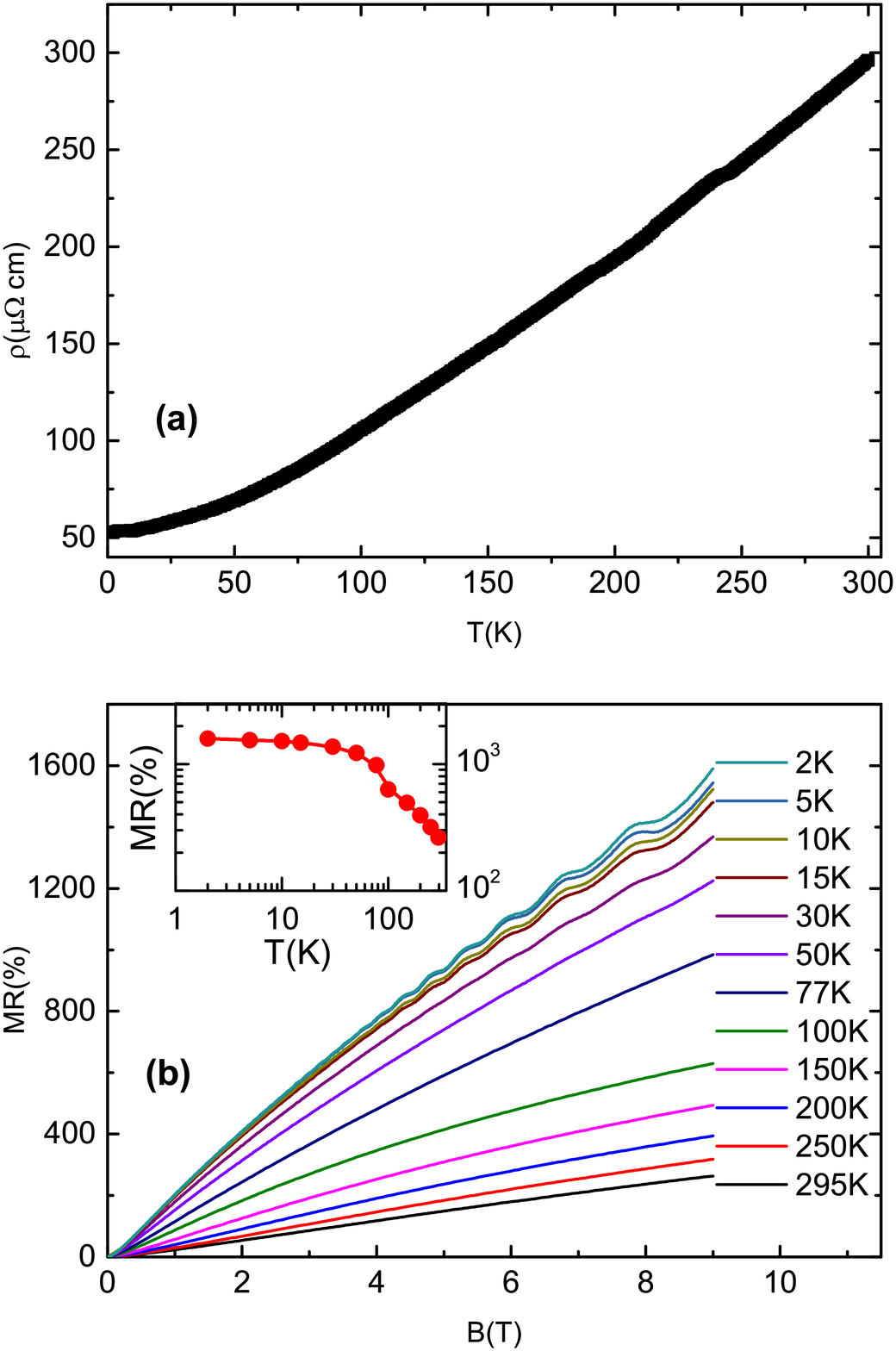}
\caption{(a) Temperature dependence of resistivity of Cd$_{3}$As$_{2}$ crystal.  (b) Magnetic field dependence of magnetoresistance (MR)  at various temperatures when field along [100] direction.(Inset temperature dependence of MR at field 9 T.)}\label{Fig.1}
\end{figure}
In these systems, the pair of 3D Dirac points is protected by crystal symmetry \cite{young,man,wang}. From Dirac semimetal, it is possible to obtain a Weyl semimetal or topological insulating phase by breaking time reversal or inversion symmetry \cite{wang1}. One can also obtain topological superconducting phase by doping charge carrier in the system \cite{wang1}. Recent experiments have demonstrated the existence of a 3D Dirac semimetal phase in Na$_3$Bi and Cd$_{3}$As$_{2}$ \cite{liu,xu,liu1,neu,bori}. Angle-resolved photoemission spectroscopy (ARPES) measurements on Cd$_{3}$As$_{2}$ single crystals have shown that two almost identical ellipsoidal  Fermi surfaces are located on the opposite sides of the Brillouin zone center point along the $k_z$ direction \cite{neu,bori}. However, these two are expected to merge into a single ellipsoidal contour around $\Gamma$ point beyond Lifshitz transition from STM \cite{jeon}. The Lifshitz transition occurs when the two Fermi surfaces touche with each other. Band structure calculations suggest that this transition should occur very close to the Fermi energy$\sim$133 meV \cite{feng}. Analysis of Shubnikov-de Haas (SdH) oscillations also suggests the existence of one or equivalent Fermi surface cross-sections corresponds to the single frequency of oscillation \cite{lphe}. Thus,  probing the Fermi surface of  Cd$_{3}$As$_{2}$ by different techniques may provide a crucial cross reference on its novel quantum properties. \\

Shubnikov-de Haas (SdH) oscillations, which have been observed \cite{lphe,andu,jeon,tian,feng} in Cd$_{3}$As$_{2}$, originate due to the oscillations in the scattering rate of charge carriers. On the other hand, the oscillations in free energy in presence of magnetic field cause de Haas van Alphen (dHvA) effect, i.e., the oscillations in magnetization. Though the oscillations in both the cases are driven by the same physical parameter, dHvA offers a more accurate picture of the Fermi surface as it is not sensitive to quantum interference effects or noise from the electrical contacts. However, the investigation of Fermi surface in this 3D Dirac semimetal through dHvA is yet to be done. In this Letter, along with SdH oscillation and high non-saturating linear magnetoresistance (MR), we report the observation of two different Fermi surface cross-sections from the dHvA effect which eventually establishes itself as a better probe to study the Fermi surface properties. Furthermore, this non-transport dHvA effect also confirms the nontrivial Berry's phase (signature of 3D Dirac fermion) in  Cd$_{3}$As$_{2}$. \\

Single crystals of Cd$_{3}$As$_{2}$ were synthesized by chemical vapor transport technique. First, we prepare polycrystalline  Cd$_{3}$As$_{2}$ samples by heating the stoichiometric mixture of high purity Cd pieces and As powder at 500$^{\circ}$C for 8 h and at 850$^{\circ}$C for 24 h in a vacuum sealed quartz tube of 15 cm long and 16 mm diameter. The product was then heated again at 550$^{\circ}$C for another 48 h for homogenization. Finally, the quartz tube was placed in a gradient furnace and heated for 48 h. During heating, the end of the quartz tube which contains the compound was maintained at 690$^{\circ}$C  while the other end was kept at  600$^{\circ}$C. The furnace was then cooled slowly to room temperature. Several small size shiny plate-like crystals formed at the cold end of the tube were mechanically extracted for transport and magnetic measurements. Powder X-ray diffraction of crashed single crystals shows that these crystals have I$_{41}$/acd space group and contain no impurity phases. The details are given in supplementary material \cite{sup}. The electrical and thermal transport measurements were carried out in a 9 T physical property measurement system (Quantum Design). The magnetization was measured using a SQUID VSM (Quantum Design). \\

\begin{figure}
\includegraphics[width=0.5\textwidth]{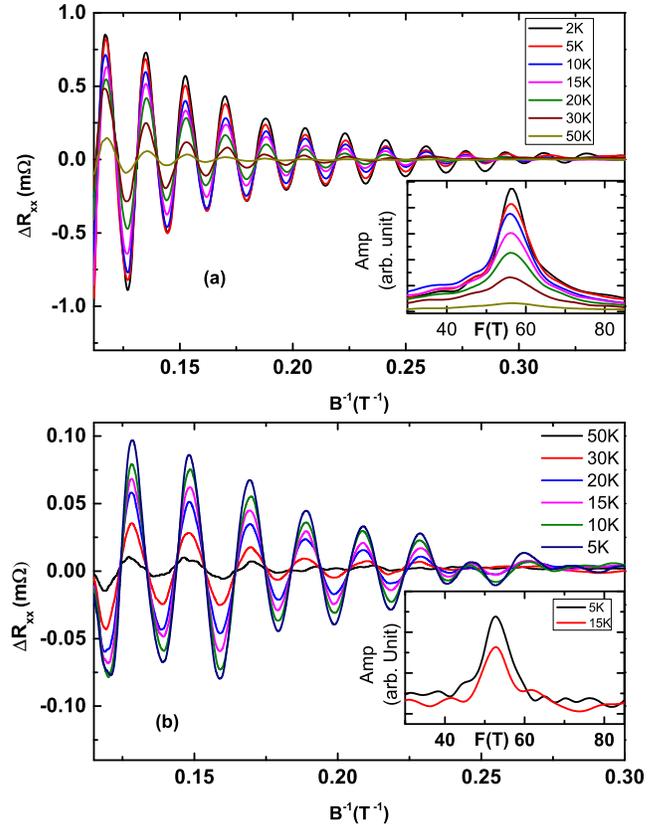}
\caption{(a) The oscillatory component $\Delta$R$_{xx}$ of MR($B$ along [100] direction) [after subtracting a smooth background] as a function of  1/$B$ at various temperature and the inset shows the oscillation frequency after fast Fourier transform. (b) $\Delta$R$_{xx}$ for magnetic field along [02\={1}] and the inset shows the corresponding frequency.}\label{Fig.2}
\end{figure}

Figure 1(a) shows the  temperature dependence of the resistivity $\rho$$_{xx}$  of a Cd$_{3}$As$_{2}$ single crystal. $\rho$$_{xx}$ is metallic over the whole range of temperature. $\rho$$_{xx}$ exhibits approximately $T^{1.5}$ dependence  above 100 K and becomes almost $T$ independent for $T$$<$20 K which reveals a residual resistivity $\rho$$_{xx}$(0) $\sim$ 50 $\mu$$\Omega$ cm.  Both the value and the nature of temperature dependence of resistivity are similar to earlier reports \cite{lphe,feng}. Fig. 1(b) displays the field dependence of the  magnetoresistance at different temperatures. MR is defined as [$\rho$$_{xx}$($B$) - $\rho$$_{xx}$(0)]/$\rho$$_{xx}$(0). Here, current is  along [012] direction and the applied magnetic field is parallel to [100] direction.  Even at room temperature and 9 T magnetic field, MR shows no sign of saturation  and its value is as high as 263$\%$. Except at low field, MR is approximately linear. Such a large linear MR at room temperature is quite unusual and needs further theoretical study for understanding the physical origin. With decreasing temperature,  MR increases rapidly and  it reaches  $\sim$1600$\%$ at 2 K and 9 T. In the inset of Fig. 1(b), we have shown the temperature dependence of MR at 9 T in log-log scale. It is clear from the figure that, unlike conventional metal, MR shows a saturation like behavior in the low temperature region below 30 K. Similar behavior has also been observed in other crystals.\\

$\Delta$$R_{xx}$, obtained after subtracting a smooth background from MR, is plotted in Figs. 2(a) and (b) as a function of 1/$B$ for fields along [100] and [02\={1}], respectively. The presence of SdH oscillations can be traced down to a field of $\sim$2 T and up to 50 K. With increasing temperature, the amplitude of oscillation decreases rapidly. The fast Fourier transform spectra of $\Delta$$R_{xx}$ versus 1/$B$ curve shows a single oscillation frequency at around 56 T [Inset of Fig. 2(a)] and 53 T [Inset of Fig. 2(b)] for field along [100] and [02\={1}], respectively. According to the Onsager relation $F$$=$($\phi$$_0$/2$\pi$$^2$)$A_F$, the cross-sectional areas of Fermi surface normal to the field are $A_F$$=$5.32$\times$10$^{-3}$ and 5.04 $\times$ 10$^{-3} {\AA}^{-2}$ respectively. Assuming a circular cross-section, a small Fermi momentum $k_F$ $\sim$0.04 ${\AA}^{-1}$  is calculated. Both the values and presence of single frequency are in good agreement with recent ARPES \cite{neu,bori} and STM \cite{jeon} results. Where in both the cases single frequency is  expected for two identical ellipsoids or single ellipsoidal contour, below and above the Lifshitz transition, respectively.\\

The magnetization of Cd$_{3}$As$_{2}$ crystal has been measured at different temperatures with $B$ parallel to  [02\={1}] direction, which displays a very clear dHvA effect [Fig. 3(a)]. We have been able to observe the oscillations down to  1 T and up to 50 K.  This  indicates that the quantization of electron orbit does not get blurred by collisions with phonon or any impurity due to the high mobility of the charge carriers. In Fig. 3(b), the magnetic susceptibility $\Delta$$\chi$ ($=$d$M$/d$B$) versus 1/$B$ plot shows how the oscillation amplitude decreases with increasing temperature. Inset of Fig. 3(b) shows the oscillation frequency obtained by the Fourier transformation of d$M$/d$B$ curve.  Unlike magnetoresistance data, we observe that there are two distinct oscillation frequencies: one at 46 T, and the other at 53 T.  This reveals  two cross-sectional areas  of the Fermi surface perpendicular to $B$,  4.39 $\times$ 10$^{-3}$ and 5.04 $\times$ 10$^{-3}$ ${\AA}^{-2}$ respectively. To understand the phenomenon, we have done the magnetization measurements with field along [100] direction \cite{sup}. But dHvA oscillation along this direction gives only one frequency which is close to the frequency found from SDH oscillation along that direction. Though the dHvA oscillation along [100] direction establishes the equivalence between two ellipsoidal Fermi surfaces found in ARPES \cite{bori} or single ellipsoidal contour  after Lifshitz transition, the magnetization measurement along [02\={1}] direction indicates two Fermi surface cross-sections correspond to the two frequencies. Also, we have done magnetization measurements in [100] plane, by applying magnetic field making an angle $\theta$ with [02\={1}] direction. It is found that the peak amplitude at 53 T diminishes with increasing $\theta$ and disappears at $\theta$ = 90$^{\circ}$ i.e., along [012] direction, giving single frequency of oscillation at 46.5 T \cite{sup}. The above study clearly shows that two equivalent cross-sections from two ellipsoidal Fermi surfaces superpose with each other  along certain direction leading to two frequencies of oscillation in dHvA effect. Another possible explanation for the occurrence of these two frequencies may be the contribution from surface and bulk states as predicted theoretically \cite{potter}. But the single frequency along [100] and [012] directions and sample thickness $\sim$0.4 mm which seems to be much larger than the critical thickness, eliminate the possibility of quantum oscillation from surface state.
\\

\begin{figure}
\includegraphics[width=0.5\textwidth]{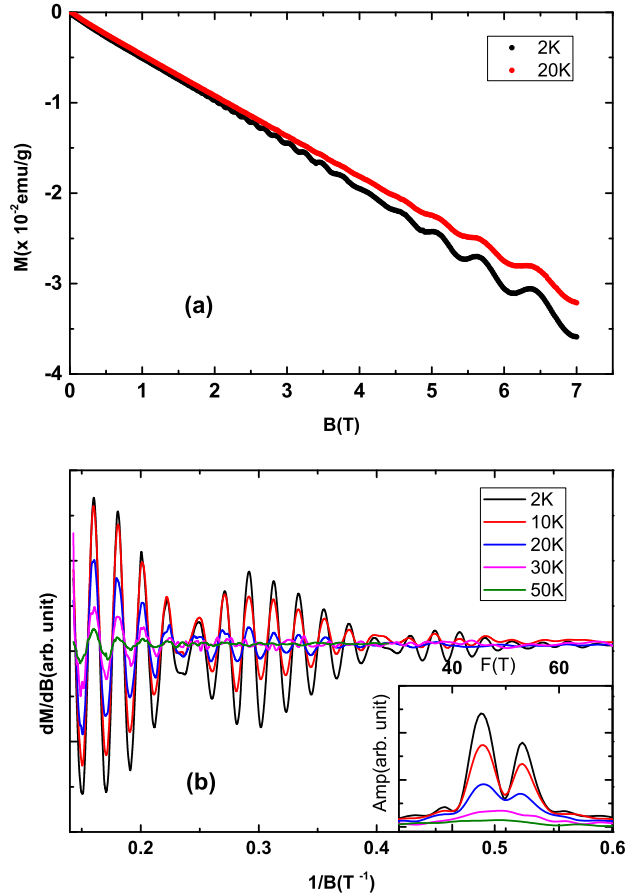}
\caption{(a) Showing magnetic field(along [02\={1}] direction) dependence of diamagnetic moment at temperature 2 K and 20 K,(b) Oscillating part of susceptibility $\Delta\chi$= d($M$)/d$B$ vs. 1/$B$ plot (de Haas van Alphen (dHvA) effect) and inset shows oscillation frequency after fast Fourier transform found from dHvA effect.}\label{Fig.3}
\end{figure}

In Figs. 4(a) and (b), the integer index $n$, corresponding to the peak in SdH and $n$+1/4 corresponding to peak in dHvA oscillation, is assigned in such a way that a linear extrapolation of the straight line yields an intercept closest to zero index. Here we have taken into account the fact that $\Delta\chi = dM/dB$ is advance in phase by $\pi$/2 with respect to $\Delta M$ (oscillation in magnetization after subtracting a smooth polynomial back ground)\cite{sup}.
In order to determine the quantum limit, i.e., the value of magnetic field required to achieve the lowest Landau level ($n$=0), we have plotted $n$ vs 1/B in Figs. 4(a) and (b). From the linear extrapolation of index plot of SdH and dHvA effect, it is found that the system reaches quantum limit at$\sim$ 50-60 T. The value of critical field for the quantum limit is very close to that observed recently \cite{narayan,cao}.
When a closed cyclotron orbit is quantized under an external magnetic field B, then according to the Lifshitz-Onsager quantization rule $A_{F}(\hbar/eB) = 2\pi(n + 1/2 -\beta+\delta)$  Here, $2\pi\beta$ is the Berry's phase where $\beta$ can take values from 0 to 1/2 (0 for parabolic dispersion as in the case of conventional metals and 1/2 for the Dirac system), and $2\pi\delta$ is an additional phase shift where $\delta$ can change from 0 for a quasi-2D cylindrical Fermi surface to ±1/8 for a corrugated 3D Fermi surface [25]. From the linear extrapolation of index plot, the intercept 0.33 (0.20) obtained in Fig. 4(a) from the SdH oscillation with B along [100] and ([02\={1}]). The intercept should be within (+,-)1/8 from zero depending on the curvature along z-direction for the 3D Dirac fermion system \cite{Mura}. But we have found an additional phase shift $\delta \sim$ 0.32 (0.2), deviating from the expected value. From the index plot of dHvA oscillation, we have found two intercepts at 0.23 and 0.35 due to the presence of two frequencies.
Again in our experiment we have used B $<$ 9 T and the Landau index plots are very straight, so one cannot expect the phase shifting due to the orbital and Zeeman splitting of the spin-degenerate conduction band \cite{jeon}. So, we assume that this extra phase shift may be due to the contribution from a small parabolic curvature and anisotropic ellipsoidal nature of Fermi surface away from the Dirac point. This result clearly demonstrates a nontrivial Berry's phase associated with 3D Dirac semimetal phase in Cd$_{3}$As$_{2}$.\\

\begin{figure}
\includegraphics[width=0.5\textwidth]{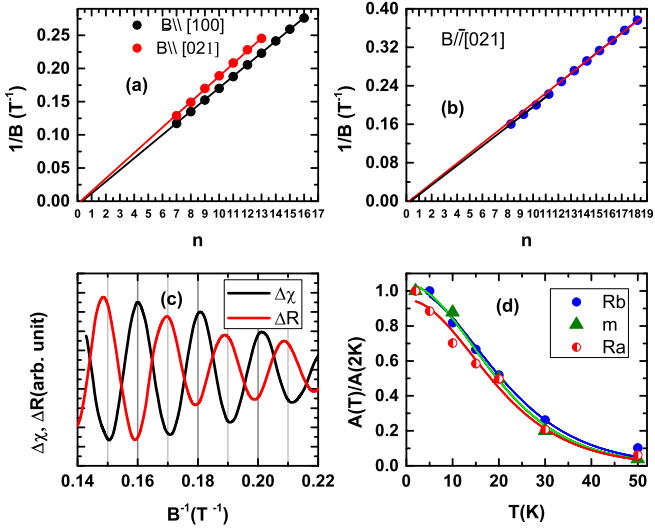}
\caption{(a) Landau index ($n$) plotted against 1/$B$ (Closed circle denote the peak position) from SdH oscillation (From the linear extrapolation of fitted line, intercept for $B$ $\parallel$ [100] is 0.32 and for $B$ $\parallel$ [02\={1}] is 0.20), (b)Landau index $n$ +1/4 plotted against 1/$B$ (Closed circle denote the peak position) from dHvA oscillation (Intercept at 0.23 and 0.35 for two parallel line) (c) Quantum oscillations of susceptibility $\Delta\chi$(1/B) and of resistance $\Delta$R(1/$B$) at 10 K temperature, (d)Temperature dependence of the relative amplitude of quantum oscillation for the 9th Landau level where $Ra$  and $Rb$  are the oscillation amplitude in SdH oscillation for B$\parallel$[100] and B$\parallel$[02\={1}] respectively and $m$ corresponds to the oscillation in susceptibility .The solid line is a fit to the Lifshitz-Kosevich formula(Equ. 1).}\label{Fig.4}
\end{figure}

Both SdH and dHvA oscillations are appropriate tools to study the Fermi surface properties and to identify the type of charge carrier. Figure 4(c) shows the measured magnetic susceptibility and the oscillating part of magnetoresistance as a function of 1/$B$. The oscillation in $\Delta$$R_{xx}$ is approximately 180$^{\circ}$ out of phase with respect to $\Delta$$\chi$. As the oscillatory component of the conductivity ($\Delta$$\sigma_{xx}$) is out of phase with $\Delta$$R_{xx}$, i.e., sign($\Delta$$\sigma_{xx}$)$=$-sign($\Delta$$R_{xx}$), both $\Delta$$\sigma_{xx}$  and $\Delta$$\chi$ are in phase with each other. Now considering the relation $\Delta$$\sigma_{xx}$$=$$\mu$$\mid$$m$$\mid$$B^2$$\Delta$$\chi$ \cite{liftz} and taking into account the fact that, in- and out-of-phase behavior correspond to electrons and holes with $\mu$= +1 and $\mu$ = -1 respectively \cite{ia},  we conclude that the majority charge carrier in this system is electron. This is consistent with negative sign of thermoelectric power (TEP) of the present samples in the temperature range 10-350 K (see supplementary material \cite{sup}) and the negative sign of reported Hall coefficient \cite{lphe}. It may be noted that  TEP exhibits a linear temperature dependence similar to that observed in graphene, which is in quantitative agreement with the semiclassical Mott relation \cite{mott}.\\

The thermal damping of the oscillation amplitude $A$ can be described by the temperature dependent part of the Lifshitz-Kosevich formula:

\begin{equation}
 A=a\frac{2\pi^2k_BT/\hbar\omega_c}{\sinh(2\pi^2k_BT/\hbar\omega_c)},
\end{equation}

where $a$ is a  temperature independent constant  and $\omega_c$ is the cyclotron frequency. The energy gap $\hbar\omega_c$ can be obtained by fitting the relative amplitude of SdH and dHvA oscillations with Eq. (1) [Fig. 4(d)]. Though for a Dirac system, its linear spectrum implies zero rest mass and gives the Landau energy level  $E_n$$=$$v_F\surd(2n\hbar eB)$, their cyclotron mass is not zero. So the quantum oscillation amplitude can indeed be described by the above expression with $\omega_{c} = eB/m^{\star}_{c}$ where $m^{\star}_{c}$ is the effective cyclotron mass of charge carrier and  the Fermi velocity $v_F$$=$$\hbar$$k_F/m^{\star}_{c}$\cite{Nov,Gus,Dong}. We deduce $v_F$$\sim$ 1.03$\times$10$^6$(0.96$\times$10$^6$) m/s and $m^{\star}$ $\sim$ 0.045m$_e$(0.05m$_e$) by analyzing the SdH oscillations with field along  [02\={1}]([100]) direction. Similarly, we have fitted the amplitude of 9th Landau Level of dHvA oscillations with Eq. (1) and found v$_{F} \sim$ 1.04$\times$ 10$^{6 }$m/s and $m^{\star}_{c}$$\sim$0.044m$_e$ for $F$=53 T. Thus, the calculated values of $v_{F}$ and $m^{\star}_{c}$ from transport and magnetization data are close to each other. These values are also close to those  reported by different groups \cite{andu,lphe,tian,feng}.Now, using the value of Fermi energy and Fermi velocity, we have deduced the Fermi energy $\sim$ 270 meV. So we are above the Lifshitz transition point reported in \cite{jeon,feng} By fitting the MR oscillation amplitude with magnetic field to $exp(-2\pi^{2}k_{B}T_{D}/\hbar\omega_{c})$, we deduce the Dingle temperature $T_{D}$=16.8 K at 2 K.\\

In conclusion, we have studied dHvA and SdH oscillations as two parallel methods to probe the Fermi surface of Cd$_3$As$_2$ and the values of Fermi wave vector, Fermi velocity, and effective cyclotron mass of charge carrier have been calculated from both the techniques. In our study, dHvA effect reveals two distinct frequencies of oscillations 46 and 53 T along [02\={1}], unlike single frequency along [100] and [012],  which confirms the existence of two Fermi surface cross-sections along certain directions. Both transport (SdH) and magnetic (dHvA) measurements support a non-trivial $\pi$ Berry's phase with a phase shift which is the signature of existence of a 3D Dirac semimetal phase in Cd$_3$As$_2$.\\

\emph{Note added in proof}. After submitting our manuscript, we have come to know the results in \cite{Zhao} where they have found double frequency from SDH oscillation measurement. They have argued that Fermi surface of  Cd$_3$As$_2$ consists  two nested ellipsoids beyond the Lifshitz saddle point.

Acknowledgement: We thank N. Khan and A. Midya for their help during measurements and useful discussions. P. Dutta would also like to thanks CSIR, India for Senior Research fellowship(File no. 09/489(0086)/2010-EMR-I).

\section{Supplementary Material for ``Probing the Fermi Surface of 3D Dirac Semimetal Cd$_{3}$As$_{2}$ through de Haas-van Alphen Technique"}
\section{Crystal Structure analysis}
\begin{figure}
\includegraphics[width=0.4\textwidth]{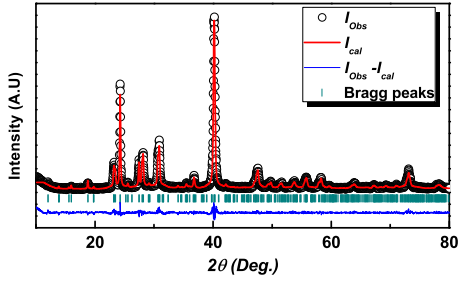}
\caption{X-Ray diffraction pattern of powdered single crystals of Cd$_{3}$As$_{2}$ . In black experimental data, in red the calculated pattern, in blue the difference and in green the Bragg positions.}\label{Fig.S1}
\end{figure}
Phase purity and the structural analysis of the samples were done by high resolution powder x-ray diffraction (XRD) technique (Rigaku, TTRAX II) using Cu-K$_{\alpha}$ radiation. Fig. S1 shows the x-ray diffraction pattern of powdered sample of Cd$_{3}$As$_{2}$  single crystals at room temperature. Within the resolution of XRD, we have not seen any peak due to impurity phase.  Using the Rietveld profile refinement program of diffraction patterns, we have calculated the lattice parameters $a$$=$$b$$=$12.64403 {\AA} and $c$$=$25.44722 {\AA} with space group symmetry I$_{41}$/acd.
\section{Magnetoresistance Measurement}
\begin{figure}
\includegraphics[width=0.4\textwidth]{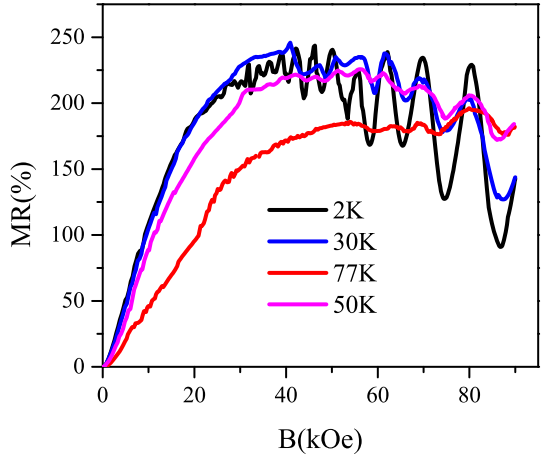}
\caption{Magnetoresistance of Cd$_{3}$As$_{2}$ single crystal with current along [012] direction and the magnetic field is applied along [100] direction.}\label{Fig.S2}
\end{figure}
Figure 2 shows the magnetic field dependence of magnetoresistance at various temperatures for another piece of Cd$_{3}$As$_{2}$  single crystal. Similar to Fig. 1(b) in the text, the magnetic field was applied along [100] direction while the current was along [012]. Though, the value of magnetoresistance is small for this sample, the amplitude of SdH oscillation is much larger and it is clearly visible even at 77 K.
\section{Thermal Transport Measurement}
\begin{figure}
\includegraphics[width=0.4\textwidth]{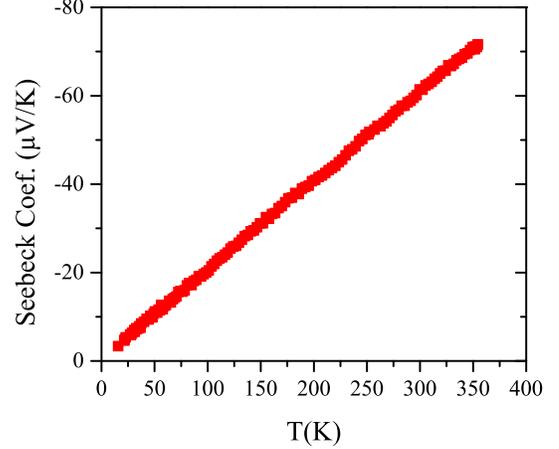}
\caption{Temperature dependence of thermoelectric power of Cd$_{3}$As$_{2}$ single crystal.}\label{Fig.S3}
\end{figure}
Both the resistivity and thermoelectric power (TEP) measurements of Cd$_{3}$As$_{2}$ single crystals were done by standard four-probe technique. Electrical contacts were made using conductive silver paste. Measurements were done in the temperature range 2-350 K in physical property measurement system (Quantum Design).  The temperature dependence of TEP is shown in Fig. 3. TEP is negative and shows linear $T$ dependence over the measured temperature range. At room temperature, the value of TEP is about 61 $\mu$V/K which is significantly large.
\section{Magnetization measurement}
\begin{figure}[h]
\includegraphics[width=0.35\textwidth]{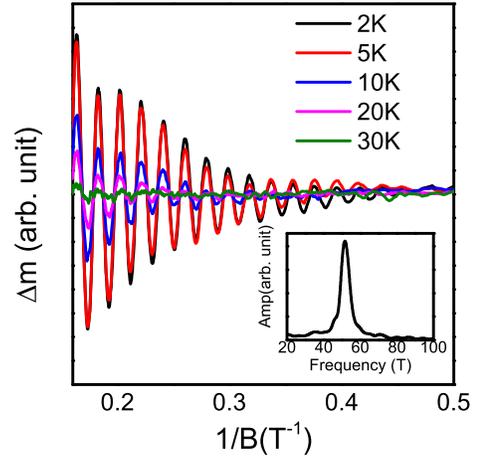}
\caption{Oscillating part of magnetization $\triangle m$ vs. 1/$B$ plot( de Haas van Alphen(dHvA) effect)for magnetic field along [100] direction and inset shows oscillation frequency at 55.20 T after Fast Fourier transform found from dHvA effect.}\label{Fig.S3}
\end{figure}
\begin{figure}[h]
\includegraphics[width=0.35\textwidth]{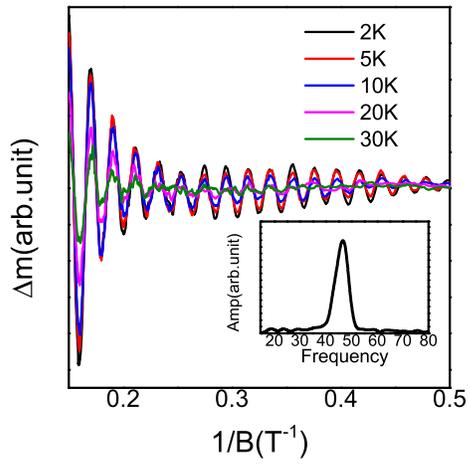}
\caption{Oscillating part of magnetization $\triangle m$ vs. 1/$B$ plot( de Haas van Alphen(dHvA) effect)for magnetic field along [012] direction and inset shows the single frequency oscillation at 46.5 T after Fast Fourier transform found from dHvA effect.}\label{Fig.S3}
\end{figure}
\end{document}